\documentclass[manuscript]{aastex}
\usepackage{natbib}
\usepackage{lscape}
\usepackage{longtable}

\shorttitle{Evolving spectral-timing properties of a type-I X-ray burst of 4U~1608--52}
\shortauthors{Chen et al.}

\begin{document}

\title{Insight-HXMT observation on 4U~1608--52: evolving spectral properties of a bright type-I X-ray burst}

\author{
Y. P. Chen$^{1}$,  S. Zhang$^{1,2}$, S. N. Zhang$^{1,2}$, L. Ji$^{3}$, L. D. Kong$^{1,2}$,
A. Santangelo$^{3}$,  J. L. Qu$^{1,2}$, F. J. Lu$^{1}$, T. P. Li$^{1,2,4}$, L. M. Song$^{1,2}$
% ...
}

\affil{$^{1}$ Key Laboratory of Particle Astrophysics, Institute of High Energy Physics, Chinese Academy of Sciences, Beijing 100049, China}
\affil{$^{2}$ University of Chinese Academy of Sciences, Chinese Academy of Sciences, Beijing 100049, China}
\affil{$^{3}$ Institut fÃŒr Astronomie und Astrophysik, Sand 1, 72076 TÃŒbingen, Germany}
\affil{$^{4}$ Department of Physics, Tsinghua University, Beijing 100084, China}
\affil{$^{5}$ Department of Astronomy, Beijing Normal University, Beijing 100088, China}
\affil{$^{6}$ Department of Engineering Physics, Tsinghua University, Beijing 100084, China}

\email{chenyp@ihep.ac.cn, szhang@ihep.ac.cn}


\begin{abstract}
The evidences  for the influence of thermonuclear (type-I) X-ray bursts  upon the surrounding environments in  neutron star low-mass X-ray binaries (LMXB) were
detected previously via spectral and timing analyses.
Benefitting from a broad energy coverage of Insight-HXMT,
we analyze one photospheric radius expansion (PRE) burst, and find  an emission excess at soft X-rays.
Our spectral analysis shows that, such an excess is not likely relevant to the disk reflection induced by the burst emission and can be attributed to an enhanced pre-burst/persistent emission. We find that the burst and enhanced persistent emissions sum up to exceed Eddington luminosity by $\sim$ 40 percentages.
We speculate that the enhanced emission is from a region  beyond the PRE radius, or through the Comptonization of  the corona.

\end{abstract}
\keywords{stars: coronae ---
stars: neutron --- X-rays: individual (4U~1608--52) --- X-rays: binaries --- X-rays: bursts}

\section{Introduction}

A low-mass X-ray binary (LMXB) hosting a neutron star (NS) occasionally undergoes outbursts with a series of  thermonuclear (type-I) X-ray bursts accompanied.
A type-I X-ray burst is caused by unstable thermonuclear burning of the accreted matter (hydrogen/helium) on the
surface of a neutron star and manifests itself as a sudden increase in the X-ray luminosity followed by an exponential decay (for reviews, see \citealp{Lewin,Cumming,Strohmayer,Galloway}). The typical duration is about tens seconds.
The most luminous bursts are the photospheric radius expansion (PRE) events,
for which the peak flux is comparable to the Eddington luminosity.

The  observations on bursts by RXTE \citep{int2013,Worpel2013,Ball2004,KeeK2014b}, NICER \citep{Keek2018,Keek2018a}, AstroSat \citep{Bhattacharyya2018} and Insight-HXMT \citep{chen2018},
revealed interactions between the burst emission and the accretion environment:  The continuum spectrum was observed to have  an enhancement at soft X-ray and/or a shortage at hard X-rays \citep{Worpel2013,Worpel2015,chen2012,ji2013}.  Such spectral deviations are considered as burst induced, and may be relevant to disk reflection and corona cooling \citep{Ball2004,KeeK2014b,Degenaar2018}.

The reflection spectrum, consisting of  discrete lines and a hump peaking at 20--40 keV,  is interpreted as disk refection of the illuminant from the corona or the boundary-layer.
The burst emission  could also serve as an illuminant to the disk.
However, so far only iron line is fairly detected during bursts, specifically during the superbursts.


4U~1608--52 is a Galactic transient LMXB  located at Galactic plane ($l=330.93^{\circ}, b=-0.85^{\circ}$). It shows outbursts at a typical frequency of once per 1--2 years.
Since its  discovery by the two Vela-5 satellites \citep{Belian},
more than 100 type-I X-ray bursts were regularly observed and
 the distance was estimated as $D\sim$2.9--4.5 kpc  (e.g. \citealp{Galloway,Poutanen}).

Moreover, some bursts from 4U~1608--52 turn out to be the brightest ones ever seen by RXTE. They reached to a flux peak of 1.2--1.5  $\times10^{-7}~{\rm erg}~{\rm cm}^{2}~{\rm s}^{-1}$,  with a burst oscillation revealed at frequency around  $\nu$=619 Hz \citep{Muno,Galloway}.
However, no pulsation is visible in the continuum emission during the outbursts.

In this study, we analyze Insight-HXMT observations of 4U~1608--52 during a soft state of its 2018 outburst by focusing on one PRE burst.
The paper is structured as that we show firstly  the  observations and data analysis,
and then give the results derived during the burst. At last, we discuss the findings in the current burst-environment interaction scenario.

\section{Observations and Data analysis}

\subsection{Swifit/XRT}
  Within the same day when Insight-HXMT detected a burst from 4U 1608--52, Swifit/XRT observed the same source.
The XRT observation was performed  $\sim$2 hours after the burst, from MJD 58304.644 to 58304.656.
The OBSID is 00010741002, with an exposure time $\sim$1 ks and a count rate $35\pm0.2$ cts/s.
The standard data process procedure of the timing mode is carried out (xrtpipeline), the spectra of source and background  are extracted with grades 0-2.
For the spectral fitting, the source spectrum is rebinned with at least 100 cts per channel.


\subsection{Insight-HXMT}
 Hard X-ray Modulation Telescope (HXMT, also
dubbed as Insight-HXMT, \citealp{Zhang2014}) excels in its broad energy band (1--250 keV)  and a large effective area in hard X-rays energy band.
It consists of three
slat-collimated instruments: the High Energy X-ray
Telescope (HE, $\sim$5000 ${\rm cm}^2$, 20--250 keV), the Medium Energy X-ray Telescope
(ME, $\sim$900 ${\rm cm}^2$, 5--30 keV), and the Low Energy X-ray Telescope (LE, $\sim$400 ${\rm cm}^2$, 1--10 keV).

  Two bursts are detected by Insight-HXMT from 4U~1608--52, as shown in Table \ref{tb}. For   the first  burst, LE do not have good exposure, we take the second burst occurred at MJD 58304.549226. As shown in Fig. \ref{fig_outburst_lc}, the contemporary count rates of MAXI and BAT are $\sim$ 75  mCrab and $\sim$ 25 mCrab respectively, suggesting that the burst is located at the soft state of the outburst.
By virtue of the quick read-out system of Insight-HXMT detectors, there is litter pile-up effect event at the PRE burst peak.
HEASOFT version 6.22.1 and Insight-HXMT Data Analysis software
(HXMTDAS) v2.01 are used to analyze the data.
Only the small field of view (FoV) mode of LE and ME is used, for preventing from the contamination of near-by sources and the bright earth.
Based on Swift/BAT source catalogue\footnote{https://swift.gsfc.nasa.gov/results/transients/}
and MAXI source list\footnote{http://maxi.riken.jp/top/slist\_ra.html}, no
bright source or active transient source  is in the  FoV of the three telescopes, as shown in the website of Insight-HXMT Bright Source Warning Tool\footnote{http://proposal.ihep.ac.cn/soft/soft2.jspx}.

The good time interval is filtered
 with the following criteria: (1) pointing offset angles
$<$0.05 degree; (2) elevation angles $>$6 degree; (3) the value
of the geomagnetic cutoff rigidity $>$6 degree.


\section{Results}
\subsection{pre-burst emission detected by Swifit/XRT and Insight-HXMT}

The non-burst/persistent emission is stable in our observations.
We jointly fitted the persistent spectra observed with XRT and HMXT, as show in Fig. \ref{sep_xrt}.
 During fittings, the LE and ME spectra are rebinned into channels having at least 100 cts, while the HE spectra are rebinned into 4 channels in 25--100 keV because of the poor statistics.
The joint fit of the spectra covers an energy band of 1--10 keV for XRT and 2--10 keV, 10--20 keV and 25--100 keV for LE, ME and HE, respectively.

We fit the Swift/XRT and Insight-HXMT(LE, ME and HE) spectra
with an absorbed
thermal Comptonization model, available as nthcomp \citep{zycki1999}
in XSPEC. The  hydrogen column (wabs in xspec)
is fixed at 1.5$\times$10$^{22}~{\rm cm}^{-2}$  \citep{Penninx1989}, to account for both the line of sight column density and as well any intrinsic absorption near the source.
In addition, a blackbody component is include corresponding to the emission from the region of the NS surface, the disk or boundary layer.
Normalization constants are included during fittings to take into account the inter-calibrations of the instruments.
 We kept the normalization factor of the Swift/XRT data with respect to the Insight/HXMT data to unity.
During the spectral analysis, a 3\% systematic error is added to account for the uncertainties of the background model and  calibration.

Using the model above, we find a acceptable  fit:$\chi_{\upsilon}$=1.08(dof 432; Fig. \ref{sep_xrt}).
The bolometric flux in 1--100 keV is $3.85\pm0.02\times10^{-9}~{\rm erg}~{\rm cm}^{-2}~{\rm s}^{-1}$, and with $\gamma\sim2.34\pm0.02$ and $T\sim0.13\pm0.02$ keV.
The constant of LE, ME and HE is 1.14$\pm$0.02, 1.17$\pm$0.02 and 1.17$\pm$0.07 respectively.

The self-consistent reflection model relxilllp \citep{Garcia}, which calculates disc reflection
features due to an illuminating power-law source,  is also tried to fit the spectra, in a way to substitute  the power-law (nthcomp) emission.
The parameters of relxilllp  are frozen to those in  \citet{Degenaar2015},  except for the  normalization $N_{\rm refl}$, the index $\Gamma$ and cut-off energy $E_{\rm cut}$
of the power law.
This results in an acceptable fit as well with a reduced: $\chi_{\upsilon}$=1.10($\upsilon$=433), although with the parameters  \citep{Degenaar2015}  derived from the hard state of 4U~1608--52.

We explore a number of different fits, e.g., absorbed diskbb+powerlaw or diskbb+nthcompt,  and find that these models can also give acceptable fit to the joint XRT and Insight-HXMT data.

\subsection{burst lightcurves by Insight-HXMT}

We show the LE/ME/HE lightcurves in Fig. \ref{fig_lc} with a time resolution of 0.25 s,
as shown in Fig. \ref{fig_lc}.
The burst profiles  exhibit a  typically fast rise and slow (exponential) decay in the soft X-ray band,
and manifest flat-topped peak (ME) and two peaks (HE) in hard X-ray band, which are typical characteristics of a PRE burst.

\subsection{broad-band spectra of burst emission by Insight-HXMT}
When we fitted the burst spectra, we estimated the background  by using the emission before the burst, i.e ., assuming the persistent emission is unchanged during the burst.
We perform the time-resolved spectroscopy with a time resolution of 0.25 s, and  define the time of the bolometric
 flux peak as a time reference
(0 point in Fig. \ref{fig_fit}).
A blackbody model (bbodyrad in Xspec) with a fixed absorption (wabs in XSPEC), as used in pre-burst fitting, is used to fit the burst spectra.
To compromise the effective area calibration deviation, a constant is added to the model.
At first attempt, for LE, the constant is fixed to 1, the others are alterable during spectra fitting. The fits indicate that most of the constant of HE are not convergent, owing to the low-significance of the HE detection. Under this situation, the constant of HE is fixed at 1 for the combined-spectra fitting.

Following the classical approach to X-ray burst spectroscopy by subtracting the persistent spectrum
and fitting the net spectrum with an absorbed blackbody.
Such a spectral model generally results in an acceptable goodness-of-fit, with a mean reduced $\chi^{2}_{\upsilon}~\sim$ 1.1.
However, we note that a significant residual is shown below 3 keV, as shown in the left panel of Fig. \ref{residual}.
Also the burst spectral fitting results
are consistent with having a detection of a PRE from the lightcurve of Insight-HXMT.

Following \citet{Worpel2013} we then include an addition component for fitting the variable persistent emission.
We assumed that during the burst the spectral shape of the persistent emission is unchanged, and only its normalization (known as a $f_{a}$ factor) is changeable.
As reported earlier by RXTE and NICER,
the $f_{a}$ model provides a better fit than the conventional one (absorbed blackbody),
We compare the above two models by using the F-test. As shown in Table \ref{table}, in some cases, the $f_{a}$  model could significantly improve the fits with a p-value $\sim10^{-5}$.

As shown in Fig. \ref{fit_1}, the spectral fitting results from these two models have differences mainly around the burst peak.  By considering an additional factor $f_{a}$, the burst blackbody flux tends to slightly decrease, and the temperature become higher but the radius shrinks.
The $f_{a}$ factor reaches a maximum of $11.64_{-2.58}^{+2.54}$, which indicates that the enhancement of the pre-burst emission is up to  $\sim$ 40\%$L_{Edd}$.    $L_{\rm Edd}$ is adopted from the flux at touch-down time (at the end of PRE phase when the radius has declined to the asymptotic value in the burst tail) during the burst.
During the PRE phase, the radius is up to $19.23_{-3.39}^{+4.56}$ km, which is three times larger than the radius measured at touch-down time  $6.20_{-0.82}^{+0.87}$ km (assuming a distance of 5 kpc). 

The soft access may  be caused by the reflection  from the accretion-disk illuminated by the blackbody emission from the NS surface/photosphere during the burst.
We employ  rdblur*bbrefl\_2xsolar\_0-5r.fits or relconv*bbrefl\_2xsolar\_0-5r.fits to replace the $f_{a}$ model, in which bbrefl\_2xsolar\_0-5r.fits and rdblur/relconv is corresponding to the reflection emission and   relativistic doppler broadening of the reflection component, respectively.

In such an attempt, the fit parameters are fixed at the dimensionless spin $\alpha$ = 0.29, the disk inclination i = 38.8 degree, the inner and outer disc radii $R_{\rm in}$=6  and $R_{\rm out}$=400 in unit of $R_{\rm ISCO}$,  which are adapted from  \citet{Degenaar2015}.
Since there are no hints of the iron emission line  or the reflection-bump in 20--40 keV, the ionization parameter  is pegged at the tabular boundary of log$\xi$ = 3.75.
However, the additional reflection component can not alleviate the residuals at soft X-rays.

\section{Discussion}
In this study, we have analyzed one PRE burst from 4U~1608--52 in the soft state of an outburst observed by Insight-HXMT.
We find a significant spectral deviation from an absorpted blackbody during the burst.
Such a deviation can be properly  accounted for by introducing an additional component denoted as $f_{a}$  that  describes the possible enhancement of the pre-burst continuum emission.
The observed enhancement evolves with the burst flux and reaches to a  peak during the expansion phase of the burst.

The enhanced continuum  emission during  bursts is  generally thought to be relevant to Poynting-Robertson drag effect,
with which the momentum of the accretion matter can be taken away and hence enlarge the local accretion rate.
Such an effect becomes even stronger for PRE.
\citet{Stahl} thought that by considering the radiation drag, the particles at the inner radius of accretion disk can move to the NS surface within $\sim 10^{-5}$ s.

Assuming  the persistent emission is proportional to the accretion rate onto the neutron star, an increase in persistent emission suggests that an additional  accretion can be induced by burst.
The increment of accretion matter induced by the burst emission,
 can be estimated as the $\sum f_{a}\dot{M}\Delta$t, where $\dot{M}$ is the pre-burst accretion rate.
This value turns out to be tiny compared with those accreted within hours or even longer before burst.

As shown in Fig. \ref{fit_1},
the $f_{a}$ value is around 10 during the radius expansion phase.
This enhancement corresponds to roughly  40\%$L_{\rm Edd}$.
Obviously, a sum up of the emission from the burst and the enhanced part from the pre-burst  gives an   overall emission level of round  140\% $L_{\rm Edd}$, as is shown in Fig. \ref{fit_ledd}.
We hence speculate  that the enhanced  persistent emission is from a region outside the burst photon spheric radius, as illuminated in Fig. \ref{illu}.
If this is the case, the inner-most disk radius  can be constrained to larger than $\sim$ 16 km.

Similar results were also reported in PRE   burst of 4U~1820--30:
 NICER observations show a photon spheric radius of $\sim$200 km and an enhanced persistent
 emission which accounts for 71\% of the overall flux and hence overwhelms the burst emission during the PRE phase \citep{Keek2018a}.
 The summed emission from the burst and the pre-burst for the enhanced part  also exceeds  $L_{\rm Edd}$.
A similar speculation is that the enhanced  persistent emissions could come from a region with a  radius of $\sim$200 km.

If there is  obscuration of the disk by the NS expanding atmosphere in the PRE phase, a decrease of  $f_{a}$ should be detected.
However, in this work, though the burst flux flatten  during the PRE phase,  $f_{a}$ and burst radius $R$ still increase and decrease  simultaneously.
\citet{Worpel2013} also found only a slight  positive correlation between the degree of  radius expansion and  the increase in persistent emission in a sample of  PRE bursts by RXTE.
This suggest that the  disk  is so far from the NS that the expanding atmosphere does not significantly obscure accretion emission;
or $f_{a}$  are likely to depend more on the burst expanding radius, because of more irradiated area of the disk by the burst.

The interaction between the burst and inhabited persistent emission was firstly studied  from the brightest and most vigorous PRE bursts with moderate/super expansions (a factor of $\sim$10-10$^{4}$ increases in emission area).
During these PRE events, the continuum emissions were missing probably due to the obscuration or hold-up for the accretion by
the pulled-up photosphere ({\citealp{Muno2000} and reference therein).

The enhancement of persistent emission during burst was firstly reported by \citet{int2013}, using the jointed Chandra and RXTE/PCA observations on a PRE burst of SAX~J1808.4--3658 in 0.5--30 keV.
They found that the pre-burst emission increased by a factor of twenty.
 More  results were reported by \citet{Worpel2013,Worpel2015} with RXTE/PCA observations in 2.5--20 keV, both for PRE and non-PRE bursts.
There they found that the enhanced pre-burst emission enhanced both during the photon spheric expansion  phase of the burst (up to a factor of eighty) and in the cooling tail.
With a larger effective area  in soft X-ray band, the persistent emission change was also detected by NICER from a single non-PRE burst of Aql~X--1 and  a PRE burst of 4U~1820--30, with $f_{a}$ measured up to 2.5 \citep{Keek2018} and 10 \citep{Keek2018a}, respectively.
Spectral residual  was also reported by AstroSAT during a burst of 4U~1728--34,
and was understood partially as reprocessed burst emission by the corona but not for the enhancement of the persistent emission \citep{Bhattacharyya2018}.

Here, a similar enhancement  pre-burst/continuum emission is detected by Insight-HXMT during a PRE burst of 4U~1608--52, with $f_{a}$ measured up to  $11.64_{-2.58}^{+2.54}$.
This is consistent with RXTE/PCA results on the bursts of 4U~1608--52 reported by \citet{ji2014},
but the broad energy coverage of Insight-HXMT allows for probing  the burst induced modification of  the persistent emission in more details.

Another interpretation of the enhanced pre-burst emission is  the  Comptonization  of the burst photon by the corona.
In principle, the burst spectrum may be influenced  due to the Comptonization of the burst photons by the surrounding corona.  As a result,
  the burst spectrum may become harder or the continuum spectrum may become softer. Both can result in spectral deviation once modelling the burst with a pure blackbody.  Actually the latter was observed as a hard X-ray deficit during burst in the low hard state of the outburst (\citealp{chen2012} and reference therein).
To study the former, a refined burst spectrum is  highly solicited which in turn requires   larger  detection area and broad band energy coverage which may be satisfied by the next generation of Chinese mission of so-called eXTP (enhanced X-ray Timing and Polarimetry mission) \citep{Zhang2019}.  Also a contemporary joint observation of the burst by NICER and Insight-HXMT is greatly  in our understanding of the burst influence upon the accretion environment.


\acknowledgements
This work is supported by the National Key R\&D Program of China (2016YFA0400800) and the National Natural Science Foundation of China under grants 11473027, 11733009, U1838201, U1838202.

\bibliographystyle{plainnat}

\begin{table}[ptbptbptb]
\begin{center}
\caption{The bursts OBSID and peak time of 4U~1608--52  detected by Insight/HXMT. }
\label{tb}
\begin{tabular}{cccccccccccccccccc}
\\\hline
 No  & obsid	& Time (MJD) \\\hline
1 &	P011476400701-20180629-01-01	 &	58298.318890	\\
2${\mathrm{*}}$		 &	P011476401301-20180705-01-01	 &	58304.549226	\\
\hline
\end{tabular}
\end{center}
\begin{list}{}{}
\item[${\mathrm{*}}$]{This work}
\end{list}
\end{table}

\begin{landscape}

\begin{table}[ptbptbptb]
\tiny
\begin{center}
\caption{The spectral fit parameters of the burst  detected by Insight/HXMT. }
 \label{table}
\begin{tabular}{cc|cccc|ccccc|c}
\hline
  & &  \multicolumn{4}{c}{ BB} & \multicolumn{5}{c}{  $f_{a}$} & \\\hline
 No & Time (s) &  $^{a}F$($10^{-8}$)  &  $T$ (keV) &  $R$ (km) &   $\chi^{2}(d.o.f)$  &
  $^{a}$F($10^{-8}$)  &  $T$ (keV) &  $R$ (km) & $f_{a}$ &   $\chi^{2}(d.o.f)$  & F-test value
  \\\hline
1 & -1.62 &  $5.04_{-0.59}^{+0.59}$ &  $3.04_{-0.15}^{+0.13}$ &  $3.71_{-0.68}^{+0.78}$ &   1.06(21) &
 $3.38_{-0.92}^{+0.91}$ &  $3.07_{-0.20}^{+0.19}$ &  $2.98_{-1.01}^{+1.13}$ &  $4.49_{-1.89}^{+1.89}$ &   0.83(20) & 1.7$\times 10^{-2}$
  \\
 2 & -1.38 &  $8.33_{-0.72}^{+0.72}$ &  $2.82_{-0.09}^{+0.08}$ &  $5.54_{-0.70}^{+0.76}$ &   0.75(37) &
 $8.39_{-0.89}^{+0.89}$ &  $2.82_{-0.09}^{+0.08}$ &  $5.56_{-0.80}^{+0.86}$ &  $-0.11_{}^{+1.00}$ &   0.77(36) & 0.90
  \\
 3 & -1.12 &  $9.37_{-0.85}^{+0.83}$ &  $2.43_{-0.12}^{+0.10}$ &  $7.94_{-1.07}^{+1.36}$ &   0.85(44) &
 $7.68_{-0.99}^{+1.00}$ &  $2.40_{-0.13}^{+0.11}$ &  $7.35_{-1.19}^{+1.46}$ &  $4.22_{-1.47}^{+1.47}$ &   0.68(43) &1.1$\times 10^{-3}$
  \\
 4 & -0.88 &  $11.87_{-0.89}^{+0.88}$ &  $2.33_{-0.09}^{+0.08}$ &  $9.72_{-1.12}^{+1.36}$ &   0.90(55) &
 $11.11_{-1.02}^{+1.03}$ &  $2.31_{-0.10}^{+0.09}$ &  $9.50_{-1.18}^{+1.41}$ &  $1.76_{-1.27}^{+1.27}$ &   0.89(54) &0.15
  \\
 5 & -0.62 &  $13.63_{-1.09}^{+1.21}$ &  $2.06_{-0.11}^{+0.12}$ &  $13.33_{-2.08}^{+2.47}$ &   0.90(60) &
 $12.70_{-1.28}^{+1.40}$ &  $2.20_{-0.14}^{+0.16}$ &  $11.21_{-2.16}^{+2.63}$ &  $6.38_{-2.00}^{+1.97}$ &   0.74(59) &5.2$\times 10^{-4}$
  \\
 6 & -0.38 &  $14.01_{-1.01}^{+1.10}$ &  $1.84_{-0.10}^{+0.11}$ &  $16.80_{-2.77}^{+3.31}$ &   1.05(66) &
 $12.97_{-1.21}^{+1.30}$ &  $2.01_{-0.13}^{+0.14}$ &  $13.56_{-2.74}^{+3.45}$ &  $7.95_{-2.16}^{+2.12}$ &   0.86(65) &1.9$\times 10^{-3}$
  \\
 7 & -0.12 &  $13.88_{-0.93}^{+1.00}$ &  $1.79_{-0.09}^{+0.10}$ &  $17.77_{-2.77}^{+3.29}$ &   1.26(69) &
 $12.19_{-1.13}^{+1.21}$ &  $1.98_{-0.12}^{+0.13}$ &  $13.56_{-2.70}^{+3.39}$ &  $11.64_{-2.58}^{+2.54}$ &   0.98(68) & 2.4$\times 10^{-5}$
  \\
 8 & 0.12 &  $14.33_{-0.87}^{+0.92}$ &  $1.69_{-0.08}^{+0.08}$ &  $20.27_{-2.88}^{+3.35}$ &   1.12(70) &
 $12.41_{-1.08}^{+1.12}$ &  $1.82_{-0.11}^{+0.12}$ &  $16.14_{-3.18}^{+3.83}$ &  $11.06_{-2.63}^{+2.62}$ &   0.88(69) & 2.9$\times 10^{-5}$
  \\
 9 & 0.38 &  $13.41_{-0.82}^{+0.86}$ &  $1.62_{-0.08}^{+0.08}$ &  $21.25_{-3.20}^{+3.73}$ &   1.08(66) &
 $12.00_{-1.03}^{+1.06}$ &  $1.73_{-0.11}^{+0.12}$ &  $17.67_{-3.78}^{+4.54}$ &  $8.27_{-2.80}^{+2.79}$ &   0.96(65) &3.6$\times 10^{-3}$
  \\
 10 & 0.62 &  $12.88_{-0.70}^{+0.72}$ &  $1.53_{-0.06}^{+0.07}$ &  $23.52_{-3.26}^{+3.78}$ &   1.12(69) &
 $11.25_{-0.85}^{+0.86}$ &  $1.61_{-0.08}^{+0.09}$ &  $19.72_{-3.77}^{+4.47}$ &  $8.71_{-2.50}^{+2.53}$ &   0.95(68) &5.2$\times 10^{-4}$
  \\
 11 & 0.88 &  $13.11_{-0.82}^{+0.87}$ &  $1.71_{-0.08}^{+0.09}$ &  $18.83_{-2.80}^{+3.26}$ &   1.19(63) &
 $11.82_{-1.01}^{+1.04}$ &  $1.82_{-0.11}^{+0.12}$ &  $15.90_{-3.22}^{+3.87}$ &  $7.27_{-2.58}^{+2.57}$ &   1.08(62) &8.7$\times 10^{-3}$
  \\
 12 & 1.12 &  $13.22_{-0.85}^{+0.91}$ &  $1.68_{-0.09}^{+0.09}$ &  $19.57_{-3.08}^{+3.65}$ &   0.96(64) &
 $12.04_{-1.02}^{+1.07}$ &  $1.81_{-0.11}^{+0.13}$ &  $16.22_{-3.38}^{+4.16}$ &  $7.56_{-2.40}^{+2.39}$ &   0.82(63) &9.3$\times 10^{-4}$
  \\
 13 & 1.38 &  $12.51_{-0.92}^{+1.00}$ &  $1.87_{-0.10}^{+0.11}$ &  $15.42_{-2.50}^{+2.97}$ &   0.90(60) &
 $11.26_{-1.12}^{+1.20}$ &  $2.01_{-0.13}^{+0.15}$ &  $12.73_{-2.63}^{+3.25}$ &  $7.28_{-2.14}^{+2.11}$ &   0.72(59) &1.8$\times 10^{-4}$
  \\
 14 & 1.62 &  $12.56_{-0.91}^{+0.99}$ &  $1.84_{-0.09}^{+0.10}$ &  $15.92_{-2.41}^{+2.85}$ &   1.13(62) &
 $11.62_{-1.05}^{+1.11}$ &  $1.93_{-0.11}^{+0.12}$ &  $13.89_{-2.59}^{+3.17}$ &  $5.66_{-2.22}^{+2.20}$ &   1.04(61) &1.5$\times 10^{-2}$
  \\
 15 & 1.88 &  $12.66_{-1.03}^{+1.12}$ &  $2.11_{-0.11}^{+0.12}$ &  $12.21_{-1.84}^{+2.21}$ &   1.19(57) &
 $10.96_{-1.20}^{+1.28}$ &  $2.27_{-0.14}^{+0.15}$ &  $9.77_{-1.79}^{+2.20}$ &  $8.85_{-2.01}^{+1.98}$ &   0.87(56) & 1.8$\times 10^{-5}$
  \\
 16 & 2.12 &  $13.52_{-1.01}^{+0.99}$ &  $2.35_{-0.09}^{+0.08}$ &  $10.18_{-1.10}^{+1.33}$ &   0.92(62) &
 $11.99_{-1.15}^{+1.16}$ &  $2.33_{-0.10}^{+0.09}$ &  $9.75_{-1.18}^{+1.38}$ &  $3.54_{-1.49}^{+1.49}$ &   0.84(61) &1.2$\times 10^{-2}$
  \\
 17 & 2.38 &  $13.23_{-0.87}^{+0.87}$ &  $2.62_{-0.07}^{+0.07}$ &  $8.08_{-0.78}^{+0.85}$ &   0.84(59) &
 $12.28_{-1.16}^{+1.16}$ &  $2.62_{-0.07}^{+0.07}$ &  $7.81_{-0.91}^{+0.97}$ &  $2.08_{-1.69}^{+1.69}$ &   0.83(58) &0.18
  \\
 18 & 2.62 &  $12.77_{-0.86}^{+0.86}$ &  $2.65_{-0.07}^{+0.06}$ &  $7.77_{-0.75}^{+0.81}$ &   0.96(58) &
 $12.27_{-1.12}^{+1.12}$ &  $2.65_{-0.07}^{+0.07}$ &  $7.62_{-0.87}^{+0.92}$ &  $0.98_{-1.39}^{+1.39}$ &   0.97(57) &0.48
  \\
 19 & 2.88 &  $13.65_{-0.86}^{+0.86}$ &  $2.90_{-0.06}^{+0.06}$ &  $6.70_{-0.67}^{+0.70}$ &   1.04(61) &
 $12.26_{-1.07}^{+1.07}$ &  $2.92_{-0.07}^{+0.07}$ &  $6.28_{-0.79}^{+0.83}$ &  $2.76_{-1.29}^{+1.29}$ &   0.98(60) &3.4$\times 10^{-2}$
  \\
 20 & 3.12 &  $11.01_{-0.80}^{+0.80}$ &  $2.76_{-0.07}^{+0.07}$ &  $6.65_{-0.68}^{+0.73}$ &   0.84(54) &
 $11.63_{-0.90}^{+0.74}$ &  $2.76_{-0.07}^{+0.07}$ &  $6.83_{-0.68}^{+0.64}$ &  $-1.00_{}^{+0.74}$ &   0.83(53) &0.17
  \\
 21 & 3.38 &  $10.14_{-1.17}^{+1.34}$ &  $2.67_{-0.16}^{+0.17}$ &  $6.79_{-1.06}^{+1.23}$ &   0.93(45) &
 $9.73_{-1.26}^{+1.41}$ &  $2.72_{-0.17}^{+0.19}$ &  $6.43_{-1.17}^{+1.39}$ &  $1.47_{-1.32}^{+1.31}$ &   0.92(44) &0.25
  \\
 22 & 3.62 &  $9.74_{-1.26}^{+1.48}$ &  $2.67_{-0.18}^{+0.19}$ &  $6.68_{-1.09}^{+1.29}$ &   1.00(39) &
 $9.91_{-1.32}^{+1.49}$ &  $2.65_{-0.18}^{+0.20}$ &  $6.85_{-1.29}^{+1.45}$ &  $-0.67_{}^{+1.41}$ &   1.02(38) &0.66
  \\
 23 & 3.88 &  $9.01_{-1.04}^{+1.20}$ &  $2.44_{-0.16}^{+0.18}$ &  $7.68_{-1.34}^{+1.64}$ &   1.14(40) &
 $8.82_{-1.10}^{+1.24}$ &  $2.47_{-0.17}^{+0.19}$ &  $7.41_{-1.45}^{+1.82}$ &  $0.85_{-1.25}^{+1.23}$ &   1.16(39) &0.53
  \\
 24 & 4.12 &  $8.71_{-0.95}^{+1.08}$ &  $2.27_{-0.14}^{+0.15}$ &  $8.72_{-1.42}^{+1.68}$ &   1.07(39) &
 $8.54_{-1.02}^{+1.13}$ &  $2.29_{-0.14}^{+0.16}$ &  $8.48_{-1.59}^{+1.91}$ &  $0.74_{-1.35}^{+1.34}$ &   1.09(38) &0.62
  \\
 25 & 4.38 &  $9.35_{-1.23}^{+1.48}$ &  $2.51_{-0.20}^{+0.22}$ &  $7.41_{-1.48}^{+1.87}$ &   1.24(39) &
 $8.78_{-1.37}^{+1.62}$ &  $2.67_{-0.23}^{+0.26}$ &  $6.35_{-1.47}^{+1.92}$ &  $3.54_{-1.54}^{+1.53}$ &   1.14(38) &3.9$\times 10^{-2}$
  \\
 26 & 4.62 &  $8.21_{-1.03}^{+1.21}$ &  $2.37_{-0.18}^{+0.20}$ &  $7.80_{-1.52}^{+1.94}$ &   1.01(33) &
 $7.88_{-1.11}^{+1.28}$ &  $2.46_{-0.20}^{+0.22}$ &  $7.09_{-1.58}^{+2.07}$ &  $1.94_{-1.31}^{+1.29}$ &   0.98(32) &0.14
  \\
 27 & 4.88 &  $8.59_{-1.22}^{+1.54}$ &  $2.35_{-0.22}^{+0.26}$ &  $8.10_{-1.98}^{+2.52}$ &   0.89(32) &
 $8.67_{-1.52}^{+1.99}$ &  $2.65_{-0.30}^{+0.37}$ &  $6.41_{-1.93}^{+2.75}$ &  $3.67_{-1.64}^{+1.60}$ &   0.76(31) &1.7$\times 10^{-2}$
  \\
 28 & 5.12 &  $7.39_{-0.98}^{+1.16}$ &  $2.34_{-0.17}^{+0.19}$ &  $7.60_{-1.46}^{+1.78}$ &   1.31(30) &
 $7.50_{-0.89}^{+1.04}$ &  $2.27_{-0.13}^{+0.22}$ &  $8.03_{-1.60}^{+1.55}$ &  $-1.00_{}^{+0.99}$ &   1.32(29) &0.39
  \\
 29 & 5.38 &  $6.45_{-0.78}^{+0.93}$ &  $1.94_{-0.17}^{+0.19}$ &  $10.26_{-2.40}^{+3.13}$ &   0.56(29) &
 $6.11_{-0.91}^{+1.04}$ &  $2.05_{-0.20}^{+0.23}$ &  $8.96_{-2.55}^{+3.53}$ &  $2.46_{-1.59}^{+1.57}$ &   0.50(28) &3.8$\times 10^{-2}$
  \\
 30 & 5.62 &  $5.53_{-0.95}^{+1.36}$ &  $2.17_{-0.27}^{+0.35}$ &  $7.56_{-2.46}^{+3.80}$ &   1.42(26) &
 $5.59_{-1.41}^{+2.01}$ &  $2.69_{-0.43}^{+0.52}$ &  $4.99_{-1.84}^{+3.16}$ &  $4.58_{-1.38}^{+1.36}$ &   1.06(25) &4.4$\times 10^{-3}$
  \\
 31 & 5.88 &  $5.01_{-0.60}^{+0.71}$ &  $1.90_{-0.18}^{+0.21}$ &  $9.42_{-2.61}^{+3.63}$ &   1.17(25) &
 $4.68_{-0.74}^{+0.85}$ &  $2.09_{-0.24}^{+0.28}$ &  $7.53_{-2.55}^{+3.90}$ &  $2.85_{-1.49}^{+1.46}$ &   1.07(24) &8.0$\times 10^{-2}$
  \\
\hline
\end{tabular}
\end{center}
\begin{list}{}{}
\item[${\mathrm{a}}$]{: The bolometric flux of the blackbody $F_{\rm bb}$ is in unit of $10^{-8}~{\rm erg/cm}^{2}/{\rm s}$.}
\end{list}
\end{table}

\end{landscape}


\clearpage

\begin{figure}[t]
\centering
   \includegraphics[angle=0, scale=0.4]{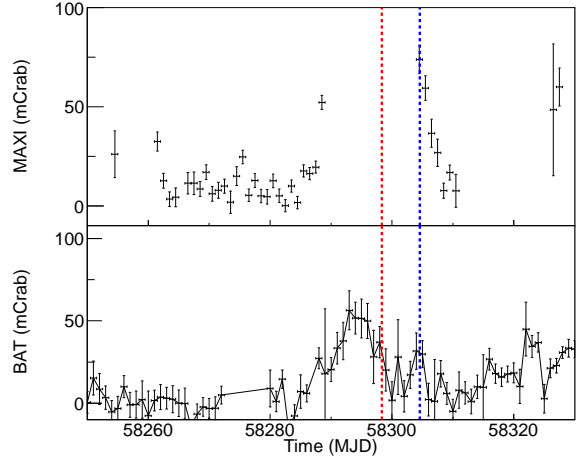}
 \caption{Daily lightcurves of 4U~1608--52 by MAXI and Swift/BAT during the outbursts in 2018, in 2--20 keV and 15--50 keV respectively. The bursts are indicated by vertical lines, the first burst and the burst of this work are marked by red and blue respectively. }
\label{fig_outburst_lc}
\end{figure}

\begin{figure}[t]
\centering
   \includegraphics[angle=270, scale=0.3]{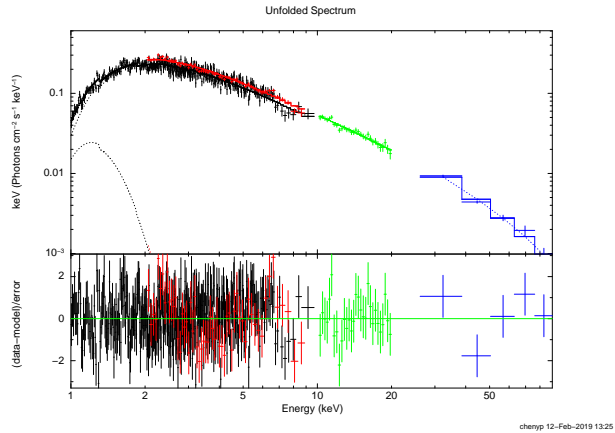}
 \caption{The spectra fit results of the  persistent emission by Swift/XRT(black), LE(red), ME(green) and HE(blue) with model cons*wabs*(blackbody+nthcomp).  }
\label{sep_xrt}
\end{figure}

\begin{figure}[t]
   \includegraphics[angle=0, scale=0.5]{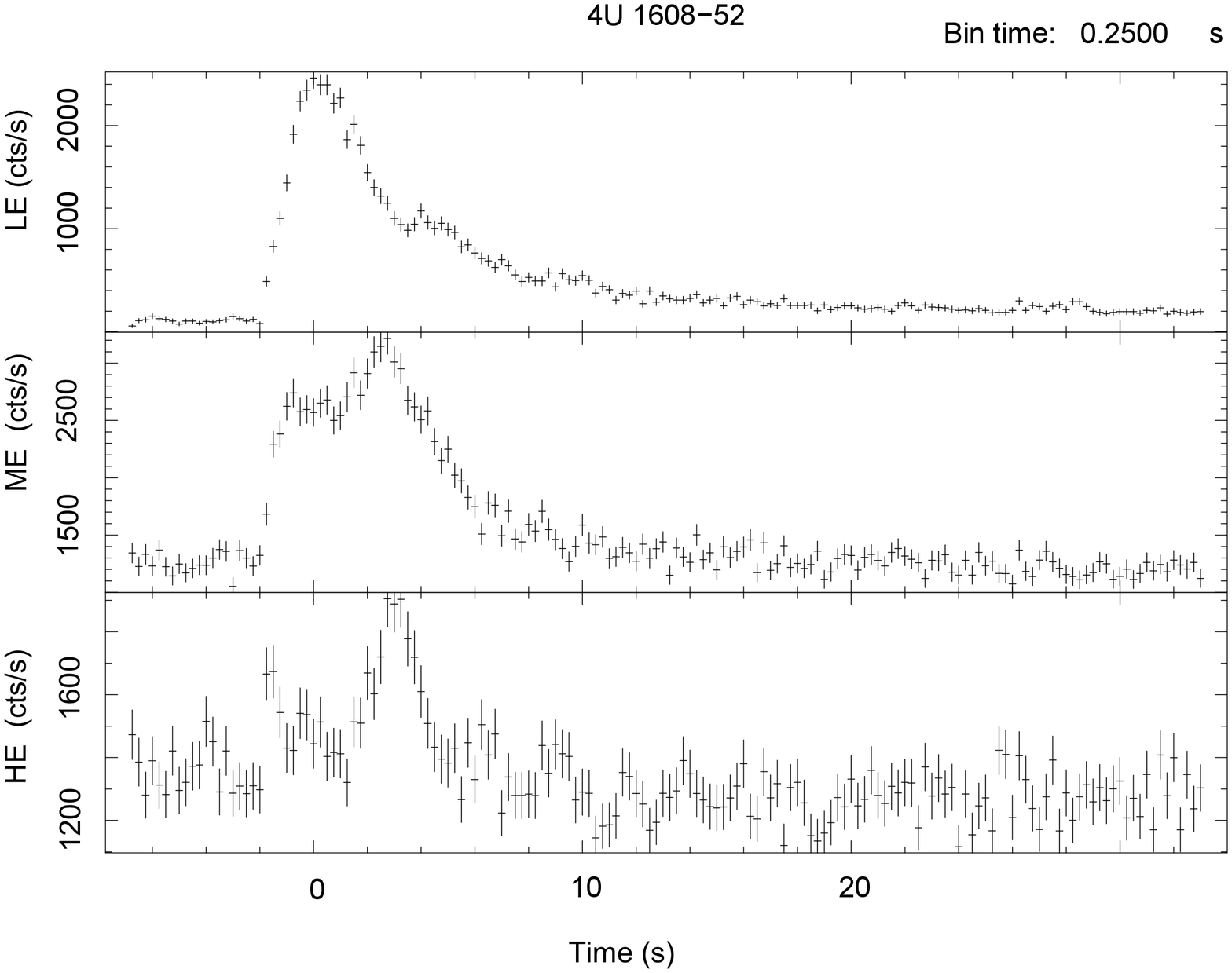}
 \caption{
 Lightcurves of the  type-I X-ray burst detected in the Insight-HXMT observation of 4U~1608--52 with time bin 0.25 s. The top, middle and bottom panel is LE, ME and HE results in their full energy band respectively. No background was subtracted.
  }
\label{fig_lc}
\end{figure}

\begin{figure}[t]
\centering
      \includegraphics[angle=0, scale=0.5]{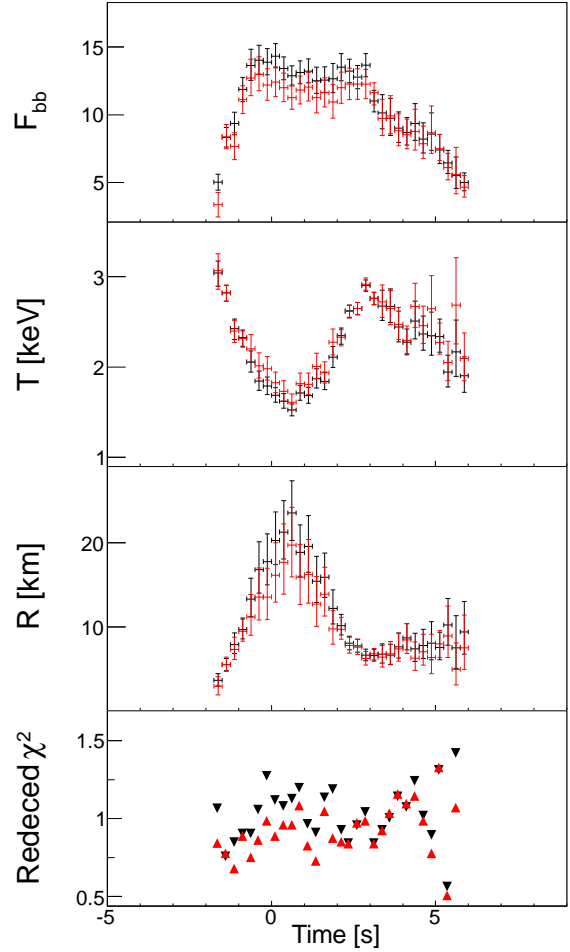}
 \caption{
Spectra fitting result of the burst with time bin 0.25 second and pre-burst spectrum subtracted,  include the time evolution of the blackbody bolometric flux, the temperature T, the observed radius of NS surface at 5 kpc, the goodness of fit $\chi_{v}^{2}$.
The black and red indicates the fitting results by a absorbed  blackbody and $f_{a}$ model, respectively.
The bolometric flux of the blackbody model $F_{\rm bb}$ is in unit of $10^{-8}~{\rm erg/cm}^{2}/{\rm s}$.  }
\label{fig_fit}
\end{figure}

\begin{figure}[t]
\centering
      \includegraphics[angle=0, scale=0.5]{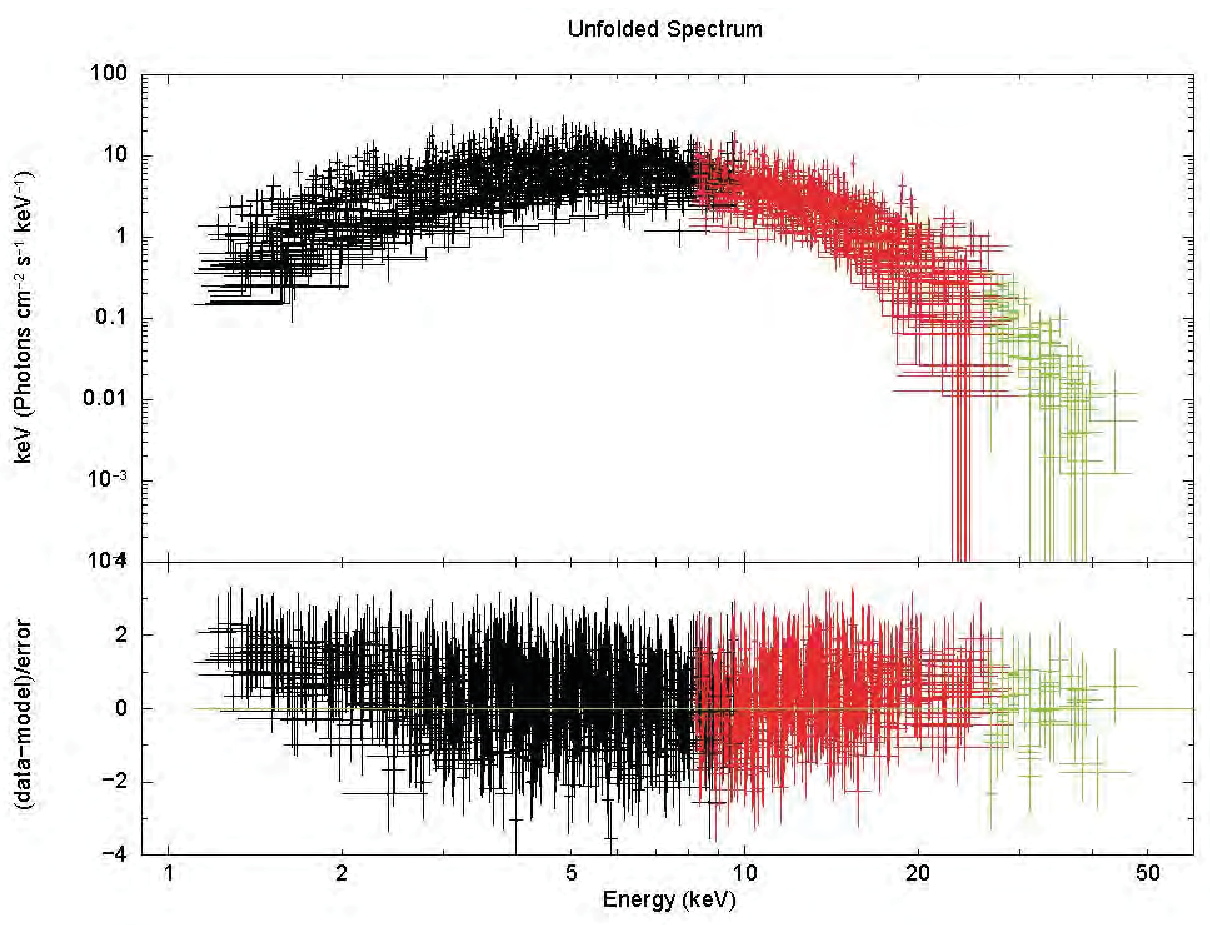}
      \includegraphics[angle=0, scale=0.5]{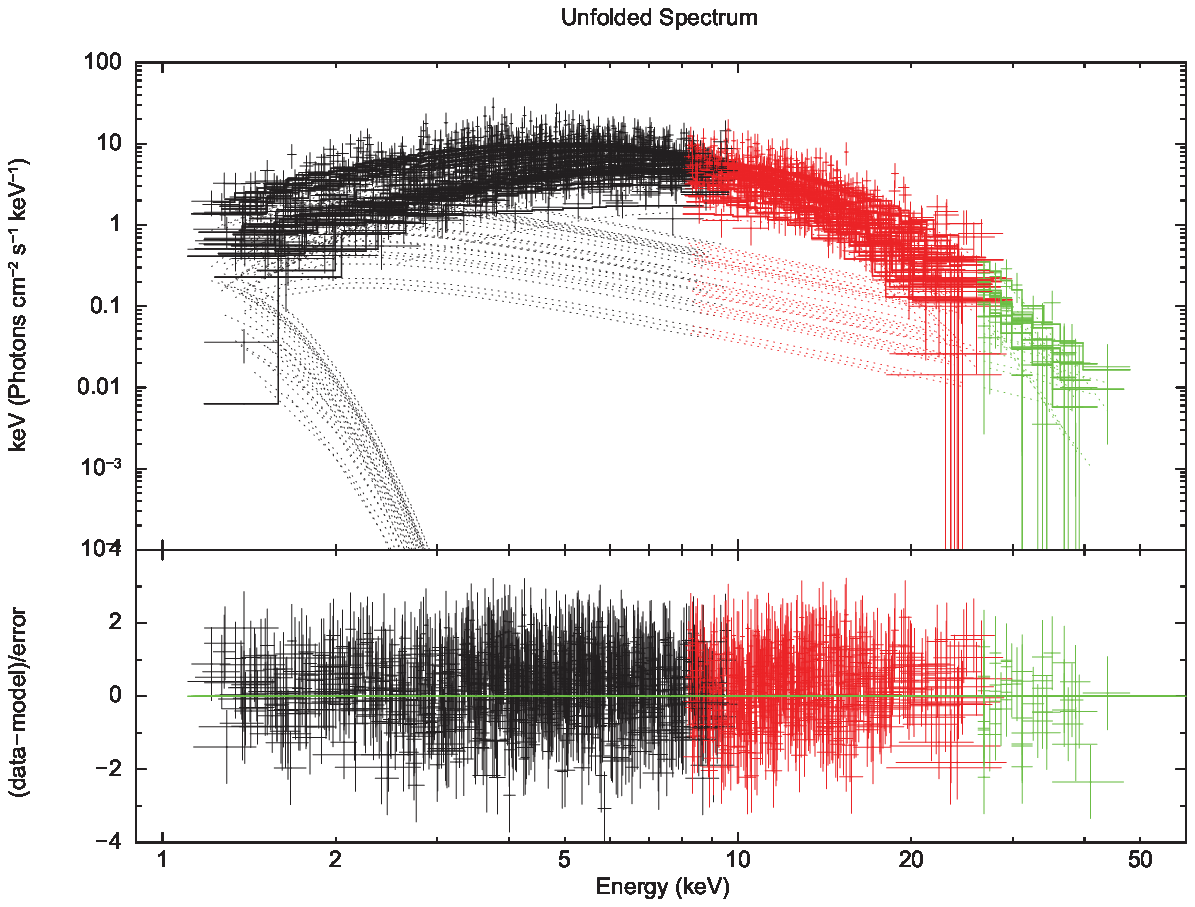}
 \caption{The merged spectral fits results by an absorbed black-body  model (left) and $f_{a}$ model (right). The lower
dotted curves represent  the model components for the pre-burst emission (absorbed bb+nthcomp).}
\label{residual}
\end{figure}

\clearpage

\begin{figure}[t]
\centering
      \includegraphics[angle=0, scale=0.4]{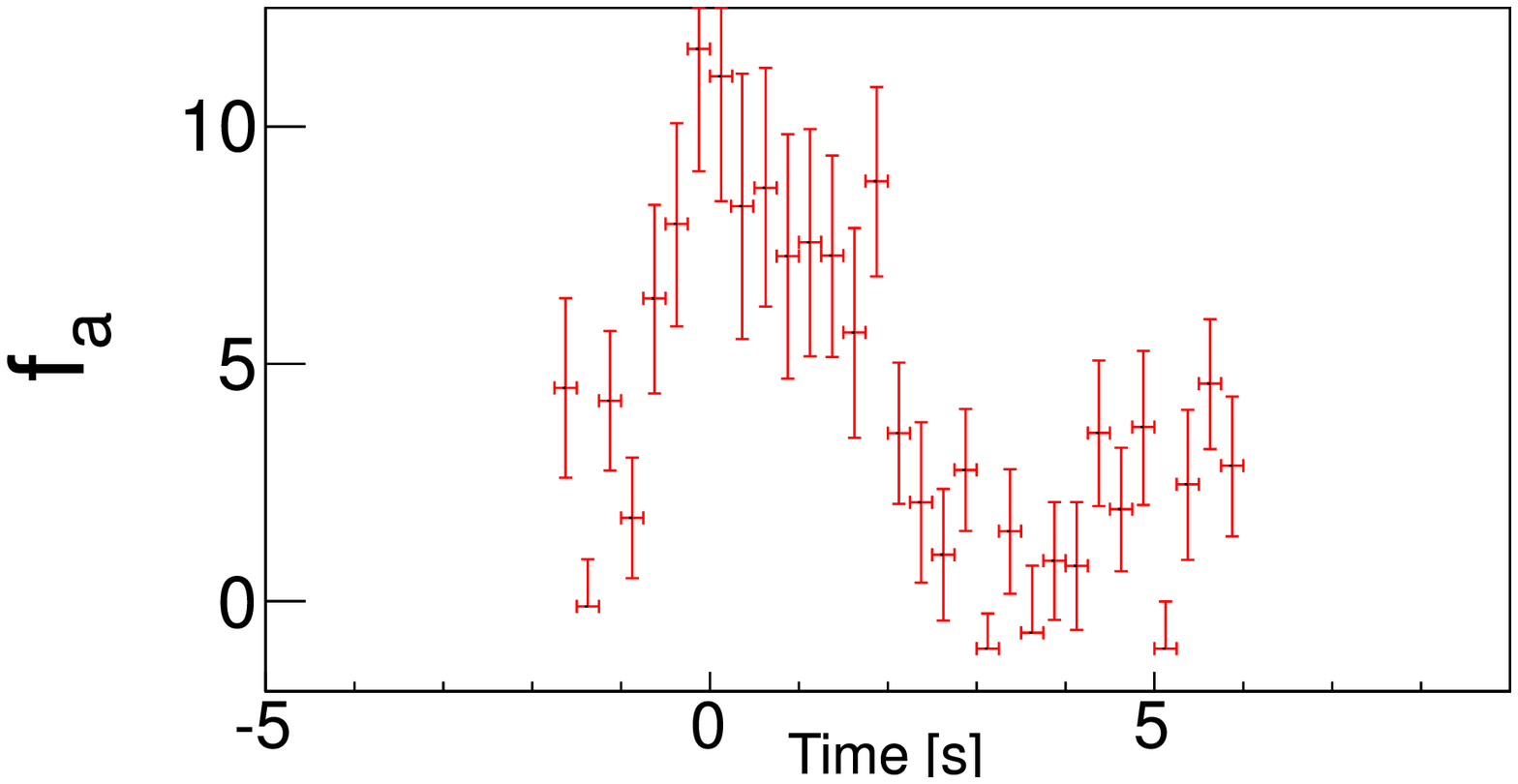}
      \includegraphics[angle=0, scale=0.4]{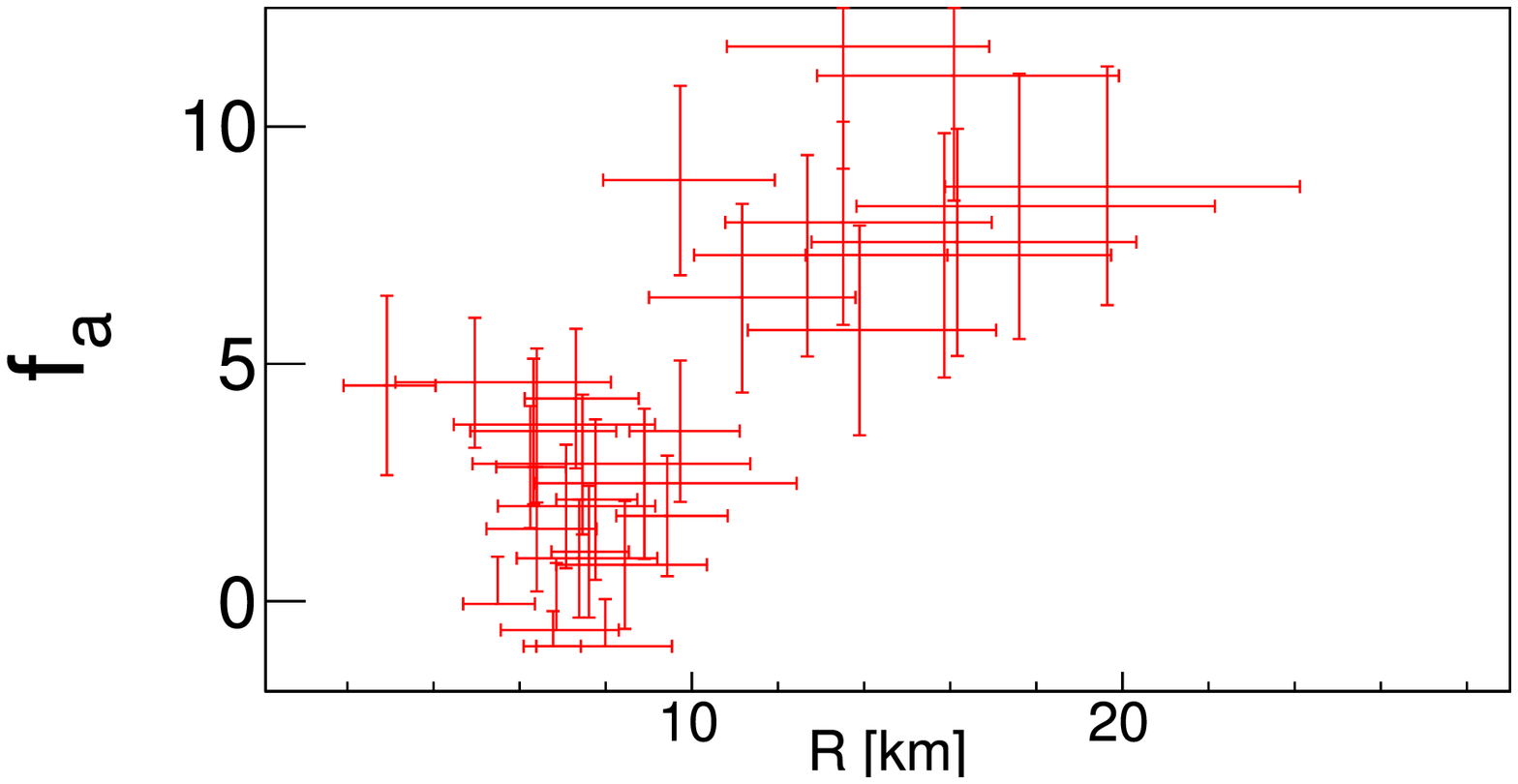}
 \caption{The  evolution (left panel) of $f_{a}$, and the plot (right  panel) between $f_{a}$ and burst radius R. }
\label{fit_1}
\end{figure}

\begin{figure}[t]
\centering
      \includegraphics[angle=0, scale=0.4]{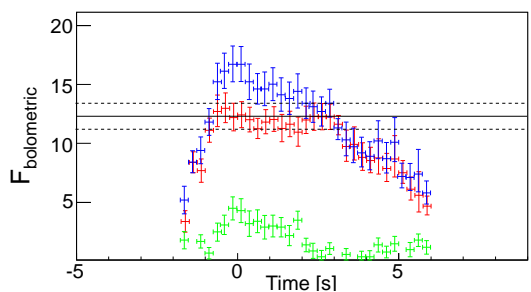}
 \caption{The bolometric flux (in unit of $10^{-8}~{\rm erg/cm}^{2}/{\rm s}$) evolution  of  the enhanced persistent emission (green), the burst emission (red) and the sum of the former two components (blue).  The horizontal solid line and dotted lines indicate the $L_{\rm Edd}$ and its upper/lower bound (1 $\sigma$), respectively.} 
\label{fit_ledd}
\end{figure}

\begin{figure}[t]
\centering
      \includegraphics[angle=0, scale=0.2]{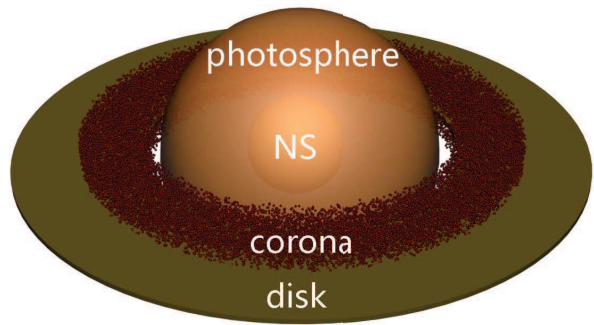}
 \caption{Illustration of the central region of an NS XRB during the PRE phase, in which the  PRE radius  is smaller than the inner-most disk radius.
} 
\label{illu}
\end{figure}

\end{document}